\documentclass[11pt]{article}
\usepackage{fleqn}
\usepackage[dvips]{graphicx}
\begin{document}

\title{Proposal to observe the strong Van der Waals force 
in $e + \bar{e} \rightarrow 2 \pi$ }

\vspace{30mm}

\author{Tetsuo Sawada\footnote{Associate member of IQS for research. \ \ \  
  e-mail address: t-sawada@fureai.or.jp }\\
Institute of Quantum Science, Nihon University, 
Tokyo, Japan 101-8308}

\date{}

\maketitle
\vspace{80mm}
\begin{flushleft}
{\large\bf Abstract}
\end{flushleft}
Large discrepancy of the p-wave phase shift data $\delta_{1}(\nu)$  
of the $\pi$-$\pi$ scattering from those of the dispersion calculation 
is pointed out. In order to determine which is correct, the pion 
form factor $F_{\pi}(\nu)$, which is the second source of information 
of the phase shift $\delta_{1}(\nu)$, is used.  It is found that the 
phase shift obtained from the dispersion is not compatible with the 
data of the pion form factor.  What is wrong with the dispersion 
calculation, is considered.

\newpage

\section{ p-wave phase shift $\delta_{1}(\nu)$ of the $\pi$-$\pi$ scattering}

  It is known that $\delta_{1}(\nu)$ is reproduced well by Wagner's 
straight line fit
\begin{equation}
\frac{s_{\rho}}{s} (\frac{\nu}{\nu_{\rho}})^{3/2} \cot \delta_{1}^{(1)}(\nu)
= \frac{s_{\rho}-s}{\sqrt{s_{\rho}} \Gamma_{\rho}}
\end{equation}
in the low energy region, where $\nu$ is the momentum squared in the center 
of mass system and $s=4 \nu +4$ in which the unit $\mu^2=1$ is adopted. 
If we compare it with the effective range function $X_{1}(\nu)$, which is 
defined by 
\begin{equation}
X_{1}(\nu)= \frac{\nu^{3/2}}{\sqrt{\nu+1}} \cot \delta_{1}^{(1)}(\nu)
\quad ,
\end{equation} 
we can rewrite Wagner's fit in terms of the effective range function  
\begin{equation}
\frac{X_{1}(\nu)}{\sqrt{\nu+1}}= \tilde{c} (\nu - \nu_{\rho})
\quad .
\end{equation} 
From the values of the mass and the width of the $\rho$-meson 
$m_{\rho}=775.65$MeV. and $\Gamma_{\rho}=143.85$MeV., the parameters 
of Eq.(3) are determined: $\nu_{\rho}=6.721$ and $\tilde{c}=-1.576$ in the 
 unit of $\mu=1$.  In figure 1, Wagner's fit and the data points are 
shown.

\begin{figure}[hptb]
\begin{minipage}{6.8cm}
\includegraphics[width=.95\textwidth,height=5.0cm]{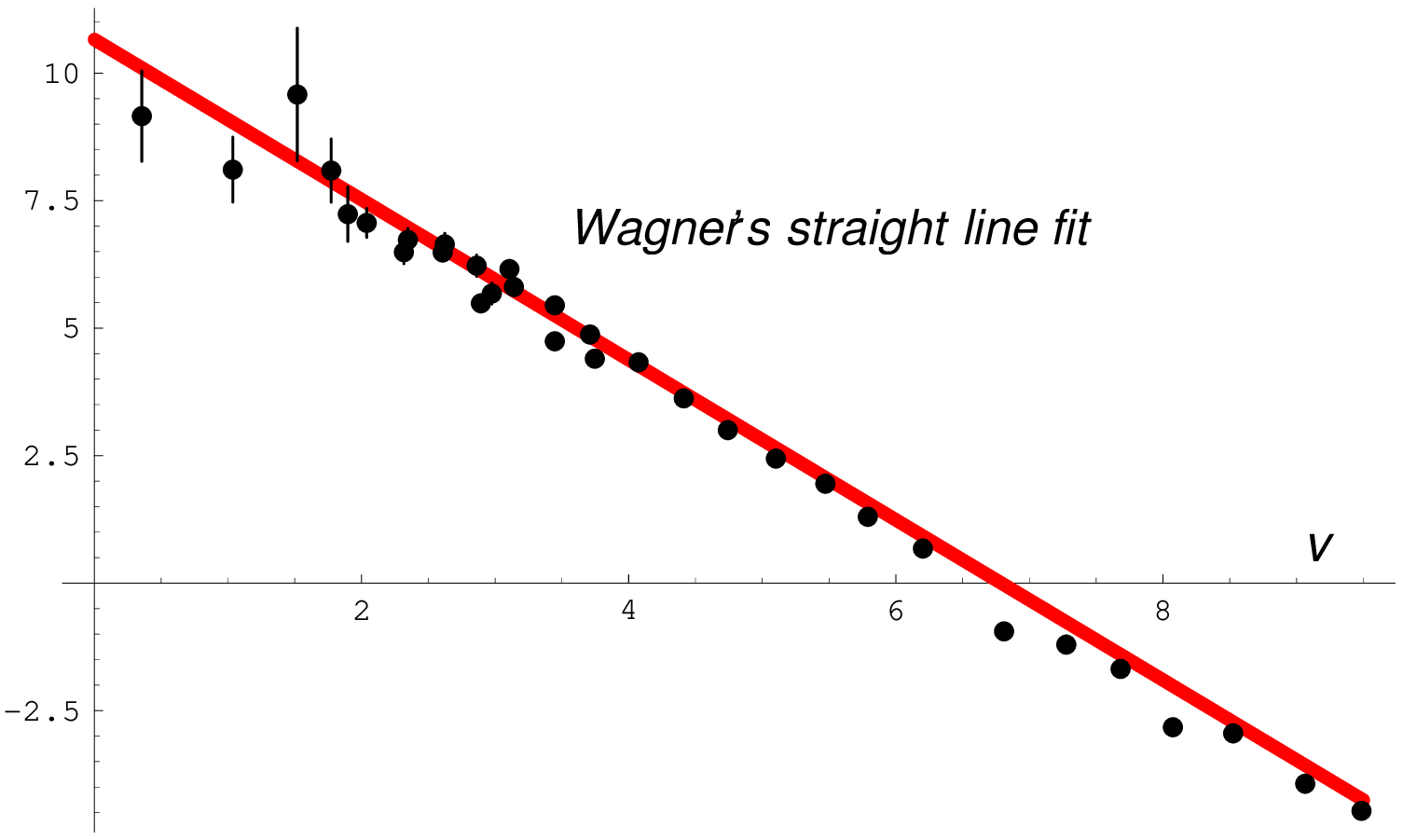}

\caption{{\footnotesize
Wagner's straight line fit of Eq.(3).   
}}
\end{minipage}
\hfill
\begin{minipage}{6.8cm}
\includegraphics[width=.95\textwidth,height=5.0cm]{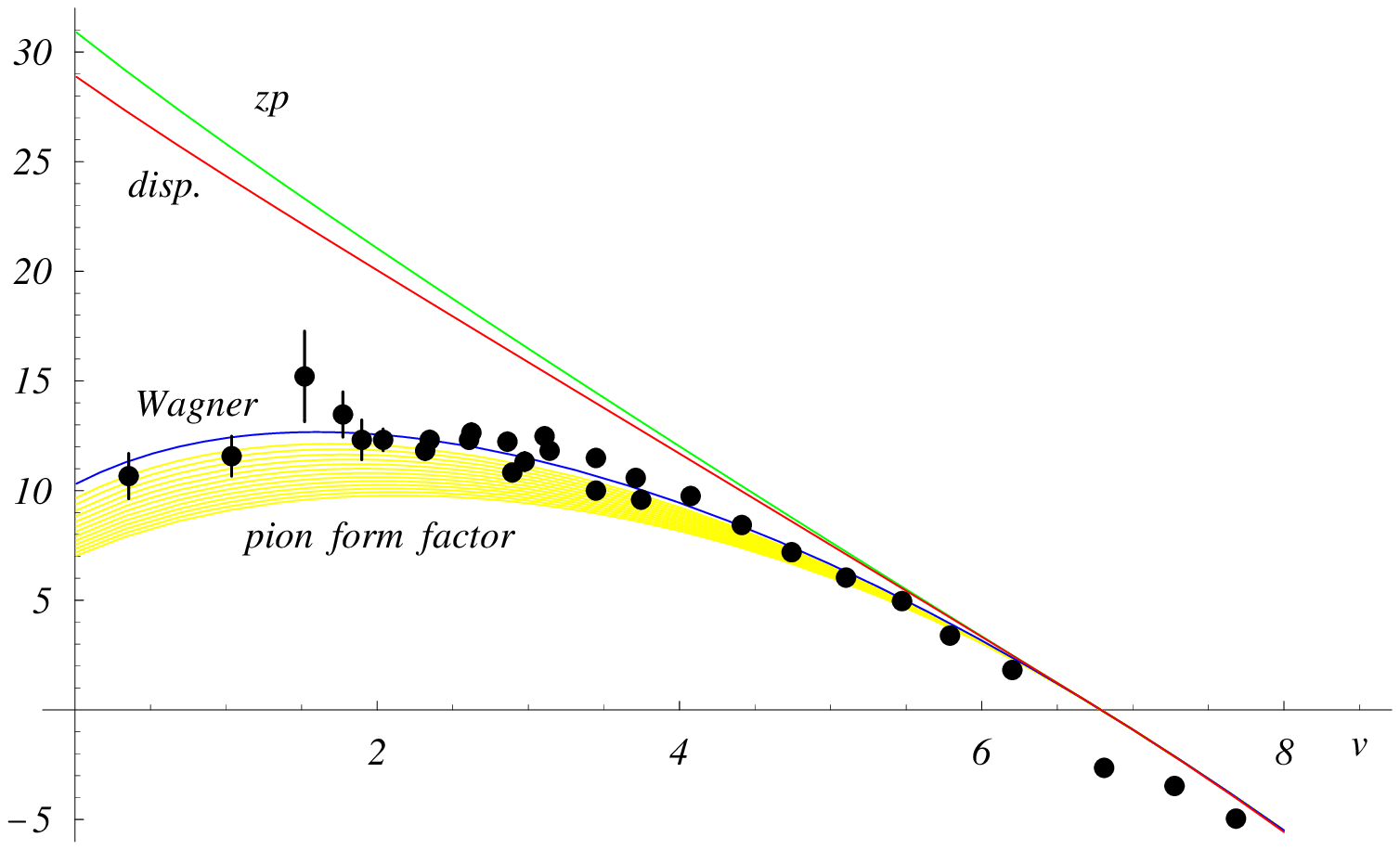}
\caption{{\footnotesize
 Effective range functions $X_{1}^{zp}(\nu)$, $X_{1}^{disp}(\nu)$ and 
 $X_{1}^{wag}(\nu).$  The corridor is the result of the pion form factor.}}
\end{minipage}
\end{figure}

     Before the dispersion calculation, it is convenient to introduce 
the zero-potential amplitude $a_{1}^{zp}(\nu)$, which is characterized 
by vanishing of the left hand spectrum.  Moreover it is expected to have 
the $\rho$-meson pole at right location which is specified by $m_{\rho}$ 
and $\Gamma_{\rho}$. The following effective range function $X_{1}^{zp}(\nu)$ 
will do the job:
\begin{eqnarray}
X_{1}^{zp}(\nu)&=& c_{0}+c_{1} \nu + \frac{2}{\pi} \frac{\nu^{3/2}}
{\sqrt{\nu+1}} \log (\sqrt{\nu}+\sqrt{\nu+1})  \\
  & & \qquad \mbox{with} \quad c_{0}=31.72 \quad \mbox{and} 
\quad c_{1}=-5.719 \quad . 
\nonumber
\end{eqnarray} 

If we remember the relation between the amplitude $a_{1}(\nu)$ and 
$X_{1}(\nu)$ 
\begin{equation}
\frac{a_{1}(\nu)}{\nu}= \frac{1}{ X_{1}(\nu)- i \nu \sqrt{\nu/(\nu+1)}}=
\frac{1}{ X_{1}(\nu)-  \nu(-\nu/(\nu+1))^{1/2}}
\end{equation} 
the necessity of the logarithmic term in Eq.(4) is evident, because the 
term $- \nu(-\nu/(\nu+1))^{1/2}$ has cuts in $\nu <-1$ as well as in 
$\nu>0$.  We can numerically confirm the property that the amplitude 
does not have the left hand spectrum by computing  
\begin{equation}
\frac{K_{1}(\nu)}{\nu} \equiv \frac{a_{1}(\nu)}{\nu} -\frac{1}{\pi} 
\int_{0}^{\infty} d \nu' \frac{ \mbox{Im}\, a_{1}(\nu')}{\nu' (\nu'-\nu)}
\quad , 
\end{equation} 
which is sometimes called Kantor amplitude.  If we use the zero-potential 
amplitude $a_{1}^{zp}(\nu)$ in evaluating Eq.(6), it must become identically 
zero, namely $K_{1}^{zp}(\nu)=0$.  In general, Kantor amplitude can be 
computed in principle from the experimental data, and can be used to 
explore the left hand spectrum namely to examine the force acting between 
the scattering particles.

    In the $\pi$-$\pi$ scattering the two-pion exchange spectrum is known to be 
computed from the crossing symmetry. The explicit form of the contribution 
of the two-pion exchange spectrum to the p-wave amplitude $a_{1}^{(1)}(\nu)/\nu$ 
is\cite{anomaly}
\begin{eqnarray}
&(& \! \! \frac{K_{1}^{2\pi}(\nu)}{\nu}\; \; ) = \\
 &=& \frac{1}{\pi}\int_{0}^{\infty} 3 \mbox{Im}
\, a_{1}^{(1)}(\nu'') \left(\frac{1+2 (\nu+1)/\nu''}{\nu^{2}} 2Q_{1}
(1+2(\nu''+1)/\nu)-\frac{1+2/\nu''}{6(\nu''+1)^{2}} \right) d\nu''  \nonumber \\
&+& \frac{1}{\pi}\int_{0}^{\infty} [ \frac{2}{3} \mbox{Im}\, a_{0}^{(0)}(\nu'')- 
\frac{5}{3} \mbox{Im}\, a_{0}^{(2)}(\nu'')]\left(
\frac{2}{\nu^{2}} Q_{1}(1+2(\nu''+1)/\nu)-\frac{1}{6(\nu''+1)^{2}} \right) d\nu'' 
\nonumber 
\end{eqnarray} 
in which Im$\, a_{\ell}^{(I)}(\nu'')$ are the imaginary part of the $\ell$-th 
partial waves of isospin $I$.  It turns out that $K_{1}^{2 \pi}
(\nu)/\nu$ is small.\cite{np2000}  The effective range function of the dispersion 
calculation $X_{1}^{disp}(\nu)$, which corresponds to the amplitude 
$(a_{1}^{zp}(\nu)+ K_{1}^{2 \pi}(\nu))/\nu$, stays close to $X_{1}^{zp}(\nu)$. 
In figure 2, the effective range curves 
 $X_{1}^{zp}(\nu)$,  $X_{1}^{disp}(\nu)$ and  $X_{1}^{wag}(\nu)$, which 
is given in Eq.(3), are shown along with the data points.
  Although the locations of the $\rho$-meson and the slopes at 
$\nu=\nu_{\rho}$ are kept the same for three curves,  $X_{1}^{wag}(\nu)$ 
deviates appreciably from other curves in the low energy region.  The 
corridor just below Wagner's curve is the effective range function obtained 
from the pion form factor in the next section.

\section{ Cross check of the p-wave phase shift $\delta_{1}(\nu)$}

   Because of the final state interaction, the phase of the pion form factor 
$F(\nu)$ coincides with the p-wave phase shift $\delta_{1}(\nu)$ at least in 
the elastic region of the corresponding $\pi$-$\pi$ scattering.  Let us 
introduce the phase function $\Delta (\nu)$ by $F_{\pi}(\nu)=|F_{\pi}(\nu)| 
e^{i \Delta(\nu)}$, which is expected to be equal to $\delta_{1}(\nu)$ in 
the low energy region.  It is covenient to define a function 
\begin{equation}
f(\nu)=\frac{1}{\nu+1} \log |F_{\pi}(\nu)|
\quad .
\end{equation} 
If we remember  
$F_{\pi}(\nu)$ is normalized at $\nu=-1$, the denominator $(\nu+1)$ in Eq.(8)  
is necessary to remove zero at $\nu=-1$ and it also serve to make $f(\nu)$ 
to decrease at large $|\nu|$. The integral representation of $f(\nu)$ has the 
form of the Hilbert transformation:
\begin{equation}
\frac{\log |F_{\pi}(\nu)|}{\nu+1}=\frac{P}{\pi} \int_{-\infty}^{\infty} 
d \nu' \theta(\nu') \frac{\Delta(\nu')}{(\nu'+1)(\nu'-\nu)}
\quad ,
\end{equation} 
and whose inversion is
\begin{equation}
\theta(\nu) \frac{\Delta(\nu)}{\nu+1}= -\frac{P}{\pi} \int_{-\infty}^{\infty} 
d \nu' \frac{\log |F_{\pi}(\nu')|}{(\nu'+1)(\nu'-\nu)}
\quad ,
\end{equation} 
because the square of the Hilbert transformation is equal to minus identity.\cite
{anomaly}

     Equations (9) and (10) enable us to extract more complete information on 
the phase shift or on the pion form factor by analyzing the phase shift and 
the form factor data jointly.  When the precise data of $|F_{\pi}(\nu)|$ were 
available in the space-like region $(\nu<-1)$ as well as in the time-like 
region $(\nu>0)$, we could use Eq.(10) to evaluate the phase $\Delta(\nu)$.
However even in such situation, we have to interpolate the data to the narrow 
unphysical region $(-1<\nu<0)$, where experimental data are not available, 
although the interpolation function is strictly restricted by the condition 
that the integration of Eq.(10) must vanish in $\nu<0$.  Since the precision 
of the data of the pion form factor in the threshold region is not sufficient, 
our program in this paper will become modest one to estimate the deviation of 
the phase shift in the sub-rho region, namely to determine the deviation 
coefficient $a$ introduced in 
\begin{equation}
\Delta(\nu)= a (\delta_{1}^{wag}(\nu)- \delta_{1}^{zp}(\nu))+
 \delta_{1}^{zp}(\nu) \quad \mbox{in} \ \  0<\nu< \nu_{\rho}
\end{equation} 
by fitting the integration of Eq.(9) to the data of the pion form factor in the 
space-like region.

\begin{figure}[hptb]
\begin{minipage}{6.8cm}
\includegraphics[width=.95\textwidth,height=5.0cm]{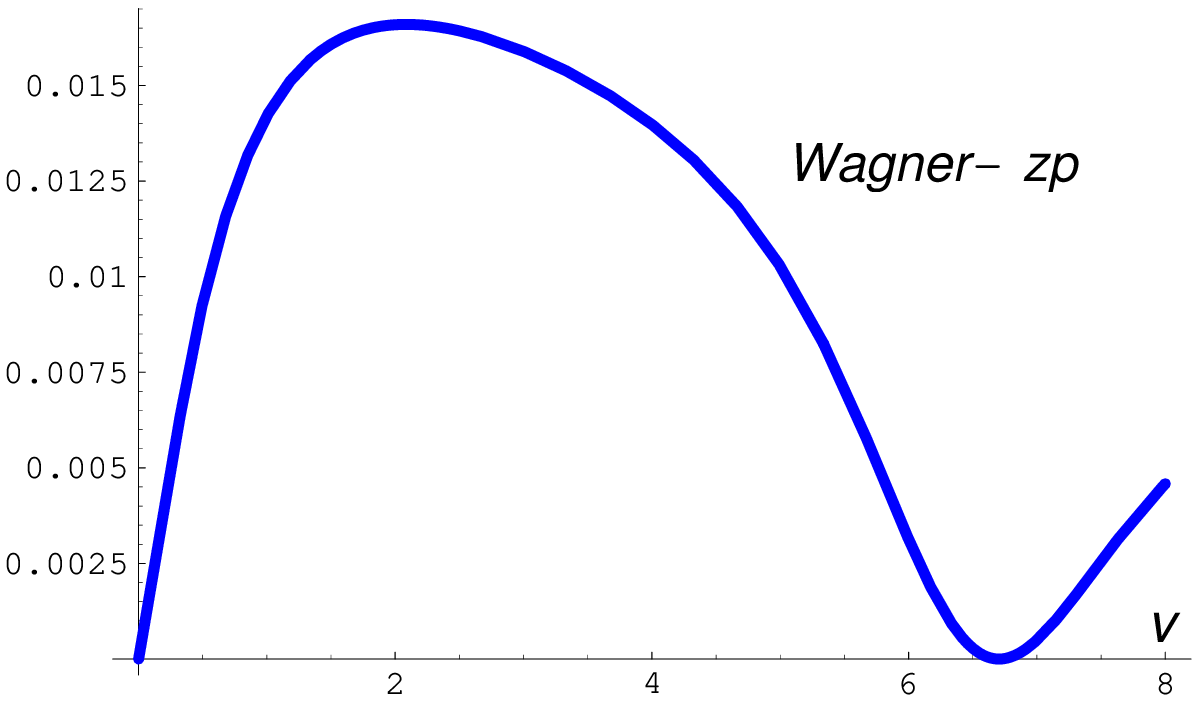}

\caption{{\footnotesize
$(\delta_{1}^{wag}(\nu)- \delta_{1}^{zp}(\nu))/(\nu'+1)$ in sub-$\rho$ region.  
}}
\end{minipage}
\hfill
\begin{minipage}{6.8cm}
\includegraphics[width=.95\textwidth,height=5.0cm]{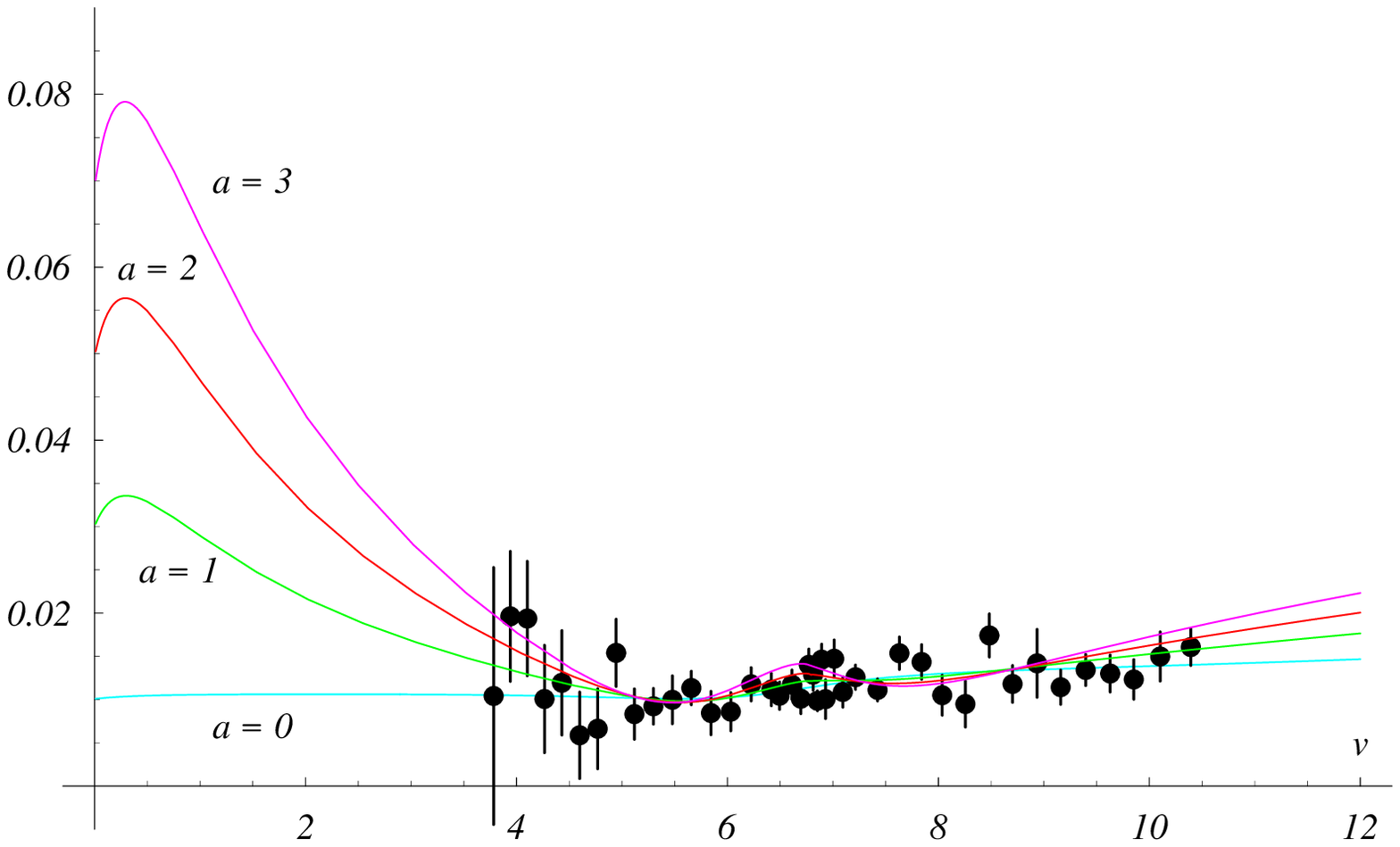}
\caption{{\footnotesize
$(f(\nu)-f^{zp}(\nu))$ in the $\rho$-resonance region for $a=$0, 1, 2 and 3.}}
\end{minipage}
\end{figure}

   In figure 3, the spectral function $(\delta_{1}^{wag}(\nu)- \delta_{1}^{zp}(\nu))
/(\nu'+1)$ necessary to calculate $(f(\nu)-f^{zp}(\nu))$ is plotted against $\nu$. 
It is interesting to examine qualitatively what is the results of the Hilbert 
transformation of this spectral function in Eq.(9).  Firstly $f(\nu)$ must shift 
upward in the space-like region, whereas in the $\rho$-meson and the higher 
energy region it must shift downward.  Secondly because of the rapid raise of the 
curve of the spectral function at small $\nu$, $f(\nu)$ must have a  
narrow peak in the threshold region.  Although the phase $\Delta(\nu)$ coincides 
with the $\pi$-$\pi$ phase shift $\delta_{1}(\nu)$ in the $\rho$-resonance 
region, they deviate each other in the higher energy region where the inelasticity 
is not negligible. Therefore before we evaluate $f(\nu)$ in the space-like 
region, we must determine $\Delta(\nu)$ in the higher energy region, for various 
values of the deviation coefficient 'a' appeared in Eq.(11), in such a way 
that it reproduces the form factor $f(\nu)$ well in the $\rho$-resonance region.
In figure 4, curves $(f(\nu)-f^{zp}(\nu))$ are plotted against $\nu$ for $a=$ 0,1, 
2 and 3, along with experimental data of CMD-2,\cite{cmd2}
 in which the $\omega$-pole is 
removed.  Although for $a=0$ the curve continues monotonously to the space-like 
region, for $a>0$ curves $(f(\nu)-f^{zp}(\nu))$ have narrow peaks in the 
threshold region.  Our proposal is to observe such a narrow peak by measuring 
the cross section of $e+\bar{e} \rightarrow 2 \pi$ precisely in the low energy 
region, and which serves to confirm that the long range force such as 
the strong Van der Waals force is acting between pions.

\begin{figure}[bhpt]
\begin{minipage}{6.8cm}
\includegraphics[width=.95\textwidth,height=5.0cm]{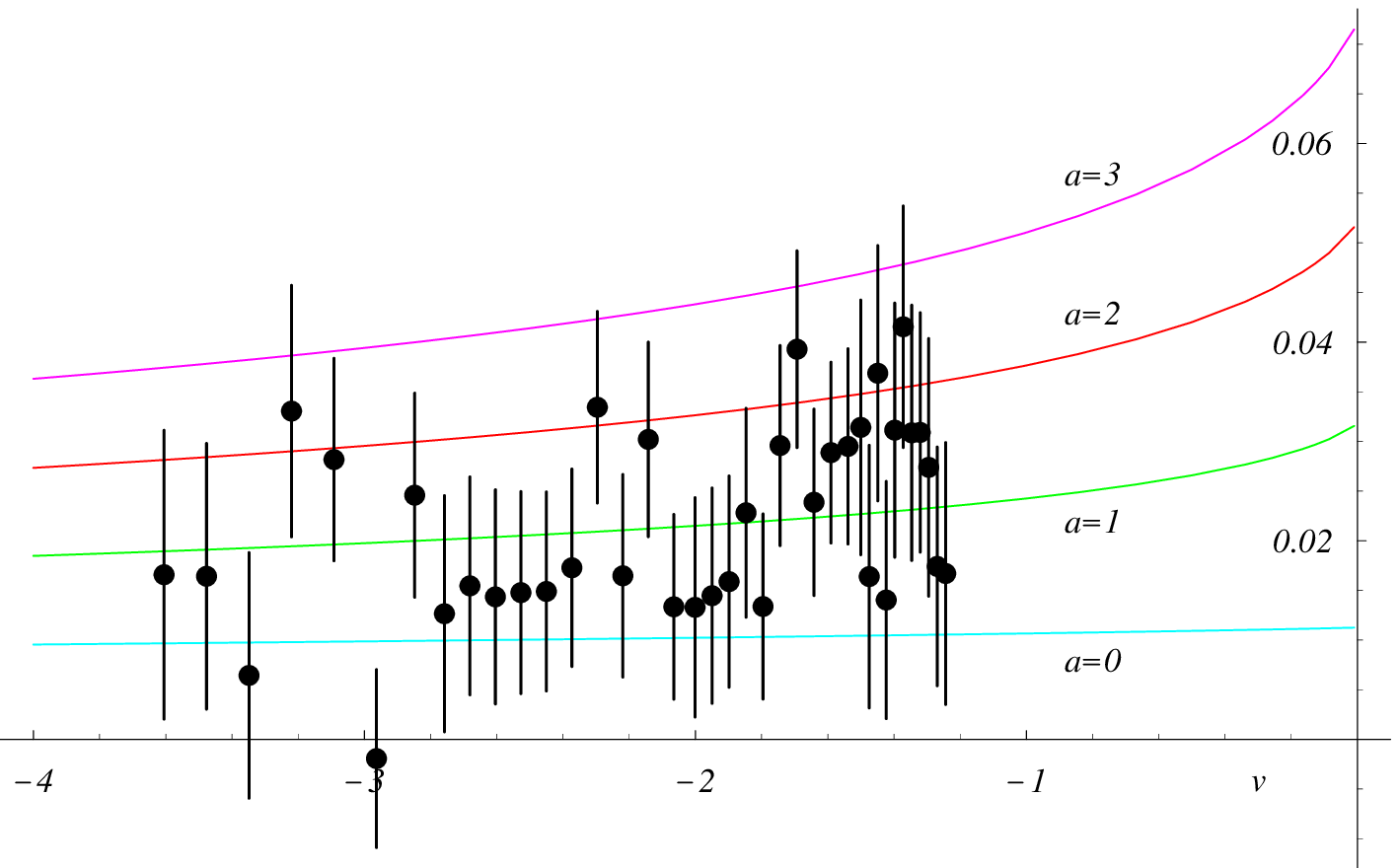}

\caption{{\footnotesize
$(f(\nu)-f^{zp}(\nu))$ in the space-like region of small momentum transfer 
 for $a=$0, 1, 2 and 3.}}

\end{minipage}
\hfill
\begin{minipage}{6.8cm}
\includegraphics[width=.95\textwidth,height=5.0cm]{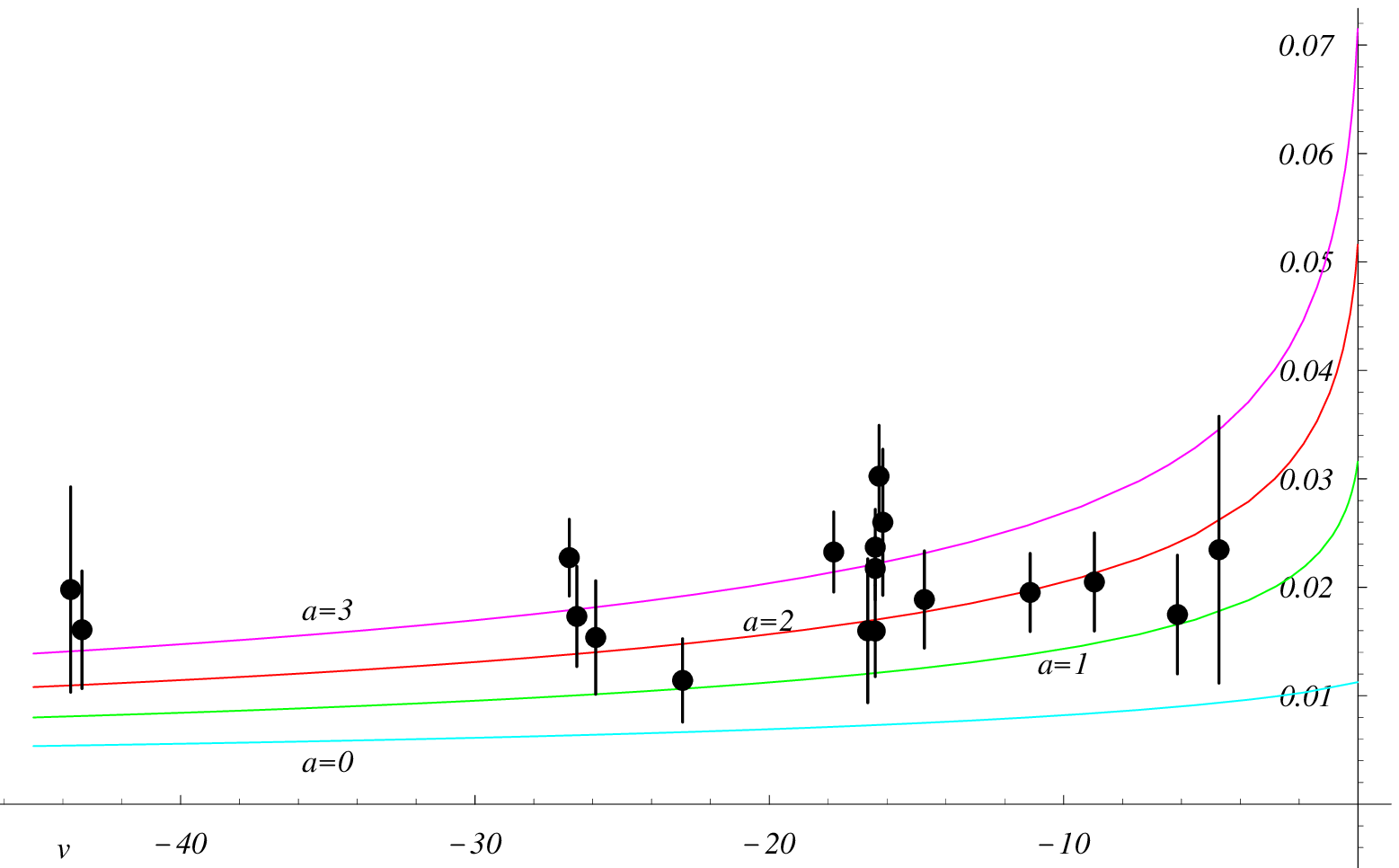}
\caption{{\footnotesize
$(f(\nu)-f^{zp}(\nu))$ in the space-like region for $a=$0, 1, 2 and 3.}}
\end{minipage}
\end{figure}

In figure 5 and 6, the same 
curves are plotted in the space-like region $(\nu<-1)$ and in the unphysical 
region $(-1<\nu<0)$.  The data points in fig.5 are those of Amendolia et.al.in 
the region of small momentum transfer, whereas the points in fig.6 are other 
data in the space-like region.\cite{spacelike}
 The graphs indicate that the data points differ from  
the curve $a=0$, which is obtained by choosing the zero-potential phase shift 
$\delta_{1}^{zp}(\nu')$ as $\Delta(\nu')$ in $0<\nu'<\nu_{\rho}$ in the evaluation 
of $f(\nu)$.   From the chi square search, $a=1.1$ is the best fit for the 
data of low momentum transfer of fig.5.   On the other hand, for the joint 
space-like data of fig.5 and 6, the minimum of chi square occurs at $a=1.7$.
Therefore the pion form factor data supports the Wagner's fit rather than the 
dispersin calculation or the zero-potential curve as shown in fig.2.

\section{ Long range interaction in the $\pi$-$\pi$ scattering}

In order to see what type of force is acting between pions, let us compute the 
contribution from the specrum on the left hand cut, namely the Kantor amplitude 
$K_{1}^{wag}(\nu)/\nu$ given in Eq.(6) by substituting $\Delta(\nu')$ by 
the phase shift $\delta_{1}^{wag} (\nu')$ which is close to the experimental data. 
In figure 7, $K_{1}^{wag}(\nu)/\nu$ is plotted against $\nu$.   

\begin{figure}[h,b,t,p]
  \includegraphics[height=.3\textheight]{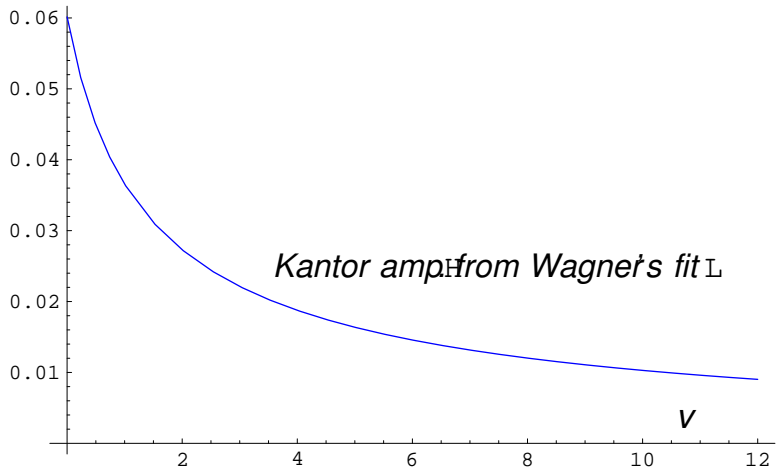}
  \caption{Kantor amplitude $K_{1}^{wag}(\nu)/\nu$  of $\pi$-$\pi$ scattering. 
A cusp of the attractive sign appears at $\nu=0$, which is characteristic to 
the long range force. 
}
\end{figure}

The curve $K_{1}^{wag}(\nu)/\nu$ is characterized by its large slope and 
very large curvature in the threshold region.  On the other hand, since the 
spectrum of the short range force starts at far left, for example the 4-pion 
exchange spectrum starts at $\nu=-4$,  $K_{1}(\nu)/\nu$ must be almost constant 
with small slope and extremely small curvature in the threshold region.  Therefore 
the curve indicates that strong force whose range is longer than that of the pion 
exchanges must acting. 

    In general the long range potential, whose asymptotic form is $V(r) \sim 
-C/r^{\alpha}$, induces a left hand spectrum in $a_{1}(\nu)$ starting from  
$\nu=0$ and the threshold behavior of the spectrum is Im$\, a_{1}(\nu')
\approx C''(-\nu')^{\gamma}$.  The powers $\alpha$ and $\gamma$ are related by 
$\alpha=2 \gamma +3$, and the coefficient $C''$ is proportional to $C$ of the 
potential.  In particular for $\alpha=6$, which is the Van der Waals potential 
of the London type, the amplitude has the singular term 
$a_{1}(\nu)=- C''\nu^{3/2}+\cdots $, whereas for $\alpha=7$, which is the Van 
der Waals potential of the Casimir-Polder type, the amplitude 
has the singular term 
$a_{1}(\nu')= C'''\nu^{2} \log \nu +\cdots $.   It is important that $C''$ and 
$C'''$ are positive for the attractive potential.   Figure 7 indicates 
that the behavior of the curve $K_{1}(\nu)/\nu$ is close to $c''_{0} -
C'' \sqrt{\nu}$, namely case of $\alpha=6$, althogh possibility of $\alpha=7$,
 namely $c'''_{0}+ C'''\nu \log \nu$, is not excluded.  We can conclude that 
the attractive Van der Waals force dominates the pion-pion interaction 
rather than the short range force.

   Finally we shall consider why the strong Van der Waals force appears in the 
hadron physics.  When the hadron was regarded as an elementary particle, the 
interaction between hadrons must occur by the exchanges of mesons, and 
therefore it was inevitably short range.  However after the introduction of the 
composite model of hadron, whose basic constructive force is strong or superstrong 
Coulomb type, because of the quantum fluctuation, we cannot avoid the strong 
Van der Waals force between the composite particles, namely between hadrons. 
Although the appearance of the Van der Waals force is simply a logical 
consequence of such composite model, what is important is its strength, which 
dominates the pion-pion interaction.  It is known that the order of magnitude of the strength $C$ of the Van der Waals potential $V(r) \sim - C/r^{6}$ is 
$C=(2/3) (^{*}e^{2})^{2} a_{1}^{2} a_{2}^{2}/\Delta E_{1}$, where $^{*}e^{2}$ 
is the "fine structure constant" of the basic Coulomb force whereas $a_{1}$ 
and $a_{2}$ are the radii of the composite particle 1 and 2 respectively. 
$\Delta E_{1}$ is the first excitation energy.  From the size of the cusp 
of figure 7, we can estimate the strength $C$, and which indicates that 
the fundamental Coulomb interaction is superstrong.  Therefore the magnetic 
monopole model of hadron must be the favorite model, because from the 
charge quantization condition of Dirac  $^{*}e^{2}$  is equal to $137/4$.
If we remember that the Van der Waals interaction is universal, we can 
expect to observe the singular behavior also in other scatterings, whenever 
sufficiently precise data are available.  In fact the attractive cusp is observed in 
the once subtracted S-wave amplitude $(a_{0}(\nu)-a_{0}(0))/\nu $ of the 
proton-proton scattering at $\nu=0$, when the repulsive Coulomb singularity 
is properly removed.\cite{nforce}

\end{document}